\documentclass{kluwer}    
\usepackage[dvips]{graphicx}
\newdisplay{guess}{Conjecture}

\begin{document}                                                                                   
\begin{article}
\begin{opening}         
\title{Understanding dwarf galaxies as galactic building blocks} 
\author{Monica \surname{Tosi}}  
\runningauthor{Monica Tosi}
\runningtitle{Dwarf galaxies as building blocks}
\institute{INAF -- Osservatorio Astronomico di Bologna}

\begin{abstract}
This is a summary of a general discussion held during the third EuroConference
on galaxy evolution. Various observational features of 
the stellar populations in present--day dwarf galaxies were
presented to introduce the discussion on the possibility that these       
systems be the main building blocks of spiral and elliptical galaxies. Many
people in the audience turned out to think that the inconsistencies among the
observed properties of large and dwarf galaxies are too many to believe that the
former are built up only by means of successive accretions of the latter.
However, theorists of hierarchical galaxy formation suggested that present--day
dwarfs are not representative of the galactic building blocks, which may be
completely invisible nowadays. Some of them suggested that, contrary to what is
usually assumed in hierarchical modelling, the actual building blocks were still
fully gaseous systems when their major mergers occurred. If this is the case,
then most of the inconsistencies can be overcome, and the scenario of
hierarchical galaxy formation becomes not too different from that of a slow     
 gas accretion.
\end{abstract}
\keywords{Dwarf galaxies, galaxy evolution, galaxy formation}

\end{opening}           

\section{Introduction} 

Observations show that galaxies merge here and there in the  local  
Universe and that big
galaxies accrete their satellites. We all know the cases of the Magellanic
Stream and Sagittarius which are being accreted by the Milky Way. Andromeda is  
quite similar in this respect, as shown by Ferguson's et al (2002). Their data  
on the red giant stars observed in M31 with the INT--WF camera indicate that  
that system contains streams and clumps just as our own Galaxy.

We learned that Cold Dark Matter (CDM) cosmology predicts that only dark
matter halos with mass smaller that 10$^8 M_{\odot}$ can form from 3~$\sigma$
fluctuations on primordial density perturbations, and that more massive systems
can only form by subsequent merging of these protogalactic fragments. Satellites
have therefore a major role and are predicted to continuously interfere with
galaxy evolution.

What we want to discuss here is whether dwarfs can be considered as the only
building blocks of massive galaxies, or their accretion is a frequent
but not necessary event. We should then discuss whether or not dwarfs properties
are consistent with a purely hierarchical scenario for galaxy formation. Another
important question is what can be the high--z counterparts of local
dwarfs, but this is not the topic of this discussion. In the following, I shall
first present a list of pro and contra arguments as discussion bases and then
briefly summarize the highlights of the actual discussion.

\section {Discussion Arguments}

\subsection {Ages}

For what concerns stellar ages, we expect the first galaxies able to form to
contain the oldest stars (see e.g. Steinmetz 2001). Early--type dwarfs clearly   
contain old stars (e.g. Horizontal Branch, HB, stars, corresponding to ages of  
12--15 
Gyr), but what about late--type dwarfs~? Well: all the examined ones have
revealed the presence of stars as old as reachable by the available photometry.
For instance, in the nearby dwarf irregular NGC 6822 the discovery of several RR
Lyraes has allowed Baldacci et al (2002) to firmly conclude that HB stars, i.e.
12--15 Gyr old stars, are present (Fig.1)

\begin{figure} 
\centerline{\includegraphics[width=20pc]{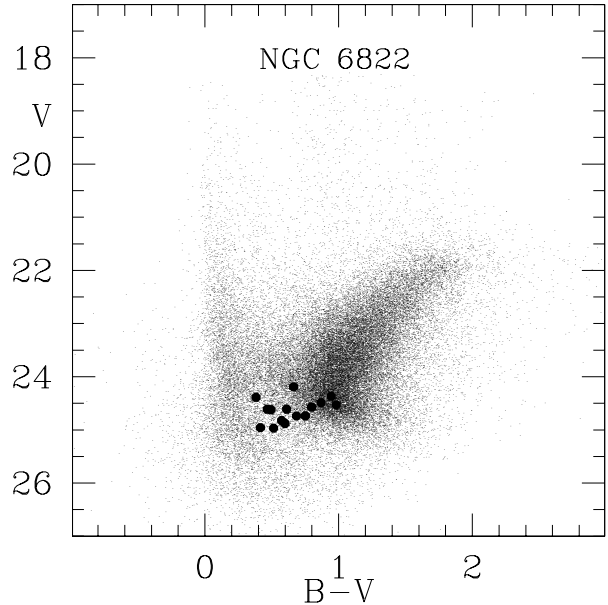}}
\caption[]{CMD of the dwarf irregular galaxy NGC 6822 as derived from VLT       
photometry. The heavier points correspond to confirmed RR Lyrae variables
(courtesy L. Baldacci).
}
\label{n6822}
\end{figure}

\begin{figure} 
\centerline{\includegraphics[width=20pc]{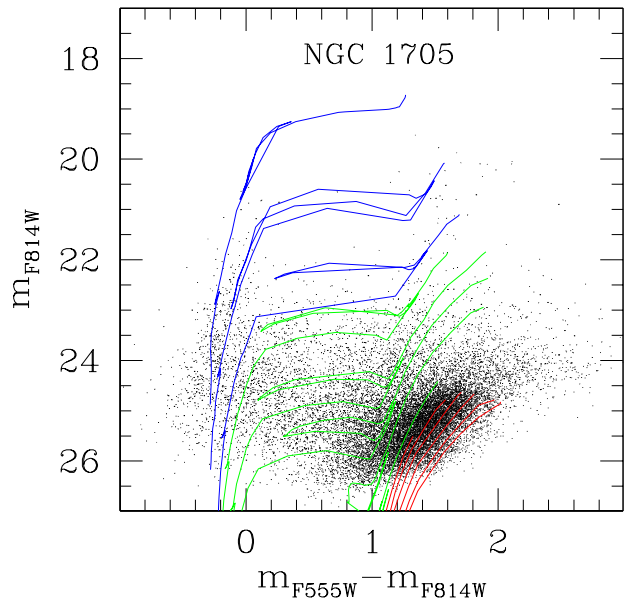}}
\caption[]{CMD of the BCD galaxy NGC 1705 as derived from HST photometry.
Overimposed are the Padova stellar evolutionary tracks from 0.6 to 30
$M_{\odot}$ (see Tosi et al 2001 for details).}
\label{n1705}
\end{figure}

In more distant galaxies, the reachable stars are obviously intrinsically 
brighter, and therefore younger. At 10--12 Mpc, even with the
 HST--WFPC2, in IZw18  
we (Aloisi et al 1999) probably reached only AGB stars, i.e. less than 1 Gyr    
old (but
the ACS should reach the tip of the red giant branch, RGB). At 5 Mpc, with the
HST--WFPC2, we (Tosi et al 2001, Annibali et al. 2002) could reach well         
below the RGB--tip (Fig.2), thus
demonstrating that the blue compact dwarf (BCD) NGC 1705, in addition to very
young stars, contains also objects at least 5 Gyr old. Table 1, an update of
one kindly provided
by U.Hopp, lists all the BCDs whose colour--magnitude diagrams (CMDs) have been
derived from HST photometry and have been interpreted in terms of stellar ages
and star formation (SF) histories. In all of them, stars as old as allowed by
the photometric magnitude limit have been found. Is this sufficient to conclude
that {\it all} dwarfs contain stars formed in the earliest Universe ? The
sample, in my opinion, is still too small to reach such a strong conclusion.

\begin{table*}
\caption[]{BCDs with SFH derived from HST CMDs}
\begin{tabular}{l|c|c|l}
\hline
Galaxy & D (Mpc) &  12+log(O/H) & Reference\\
\hline
I~Zw~18    & 10-12 &  7.18 & Aloisi et al. 1999\\
VII~Zw~403 & 4.4   &  7.69 & Lynds et al. 1998 \\
UGCA~290   & 6.7   &   ?   & Crone et al. 2002\\
I~Zw~36    & 5.8   &  7.77 & Schulte-Ladbeck et al. 2001\\
NGC~6789   & 3.6   &  7.7? & Drozdovsky et al., 2001 \\
UGC~5272   & 5.5   &  7.83 & Hopp et al., in prep\\
MrK~178    & 4.2   &  7.95 & Schulte-Ladbeck et al. 2000\\
NGC~4214   & 2.7   &  8.27 & Drozdovsky et al. 2002\\
NGC~1569   & 2.2   &  8.31 & Greggio et al. 1998\\
NGC~1705   & 5.1   &  8.36 & Annibali et al. 2002 \\
\\ \hline
\end{tabular}
\end{table*}

In any case, the mean ages of stars in ellipticals are generally older than those
in dwarfs, a feature recently confirmed by the study (Kauffmann et al 2002) of
80000 galaxies observed by the Sloan survey. An argument against pure merging
galaxy formation, since building blocks should instead
contain older stars than their daughters.

An argument which is  often presented in favour of hierarchical galaxy   
formation
is the age spread e.g. of the Milky Way halo stars: no doubt that if halo stars
have formed during its collapse and have ages differing by 1--2 Gyr, we cannot
let the halo have collapsed in 100 Myr, as in the original version of the {\it
monolithic} collapse scenario (Eggen, Lynden-Bell, Sandage 1962). But wouldn't a slower collapse be
compatible with the observed stellar ages ?

\subsection {Chemical Abundances}

We know that dwarfs (both early-- and late--type dwarfs) are metal poor
(0.0001$\leq Z \leq Z_{\odot}/2$). How about abundance ratios which are the     
signature of the SF timescales ? For instance, the [$\alpha/Fe$] ratio depends  
on the relative contributions of SNeII ($\alpha$ producers) and SNeI           
($Fe$ major producers) so that, at a given metallicity,  slow, continuous     
SFs imply
low [$\alpha/Fe$], while short SFs imply high [$\alpha/Fe$] (see Matteucci     
1992 and her figure 1 in this volume). In the few local
dwarfs where [$\alpha/Fe$] has been measured with appropriate spectroscopy, it 
ranges between the low ratio --0.2 of
the Magellanic Clouds ($\Rightarrow$ continuous SF) and the moderate ratio 0.02
$\leq$[$\alpha/Fe$]$\leq$0.13 of dSph's (Shetrone et al 2001 and Fig.3,         
Tolstoy, this volume). In the Milky Way  halo, the $\alpha$
overabundance is about 0.4--0.5; in ellipticals it is observed to range between
$\sim$0.05 and 0.25 ($\Rightarrow$ short SF episodes). How could ellipticals
(high $Z$) and halos (low $Z$) manage to get [$\alpha/Fe$]$>$0.2 if dwarfs
similar to the local ones are
their building blocks ? In fact, semi-analytical models do not reproduce
the whole range of ratios observed in ellipticals (e.g. Thomas 2001) because
they fail to get a large portion of its high [$\alpha/Fe$] end. Chemical
evolution models based on monolithic collapse do, instead, reproduce the
observed ratios (see figure 2 in Matteucci, this volume). 
If the Sf is more time-concentrated in ellipticals than in dwarfs, how can
the latter be the building blocks of the former ?

\begin{figure} 
\centerline{\includegraphics[width=26pc]{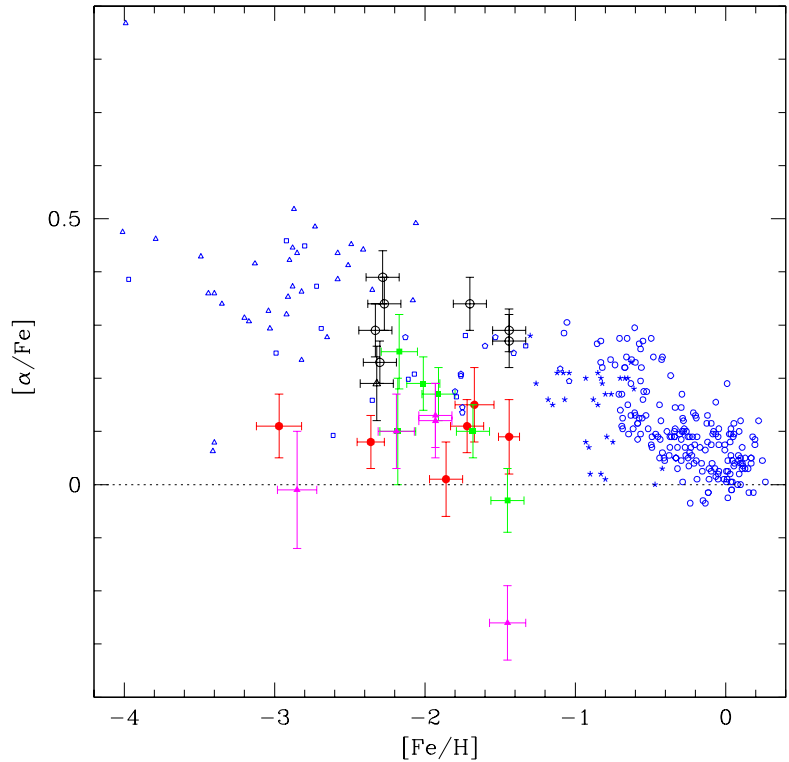}}
\caption[]{Abundance ratios derived from high-resolution spectra of Ursa Minor,
Draco and Sextans (filled dots with error bars) compared with those of Milky Way
halo and disk stars (dots) and globular clusters (open symbols with error bars). 
See Shetrone et al 2001 for details.}
\label{alpha}
\end{figure}

Metallicity provides another problem for pure hierarchical galaxy formation: the
metallicity distribution of halo globular clusters, both in the Galaxy and in
ellipticals. Actually, the multimodality of the metallicity distribution
functions of globular clusters (e.g. Harris 2001, and references therein)       
is one of the first
arguments against monolithic collapse and in favour of hierarchical formation,
buth why are the distributions most frequently bimodal, rather than multimodal,
if the accretion episodes are several tens or hundreds and at all possible
epochs ?

\subsection {Counts}

The number of dwarf satellites predicted by CDM scenarios is much larger than
observed: 11 satellites observed against 50--500 (depending on the    
assumptions)  predicted 
for the Milky Way, and 40 observed against 1000 expected DM halos in the Local
Group (see e.g. Mateo 1998, Moore et al 1999). This is the so-called            
{\it satellite catastrophe}
for hierarchical galaxy models: where have the missing satellites gone ?

\begin{figure} 
\centerline{\includegraphics[width=22pc]{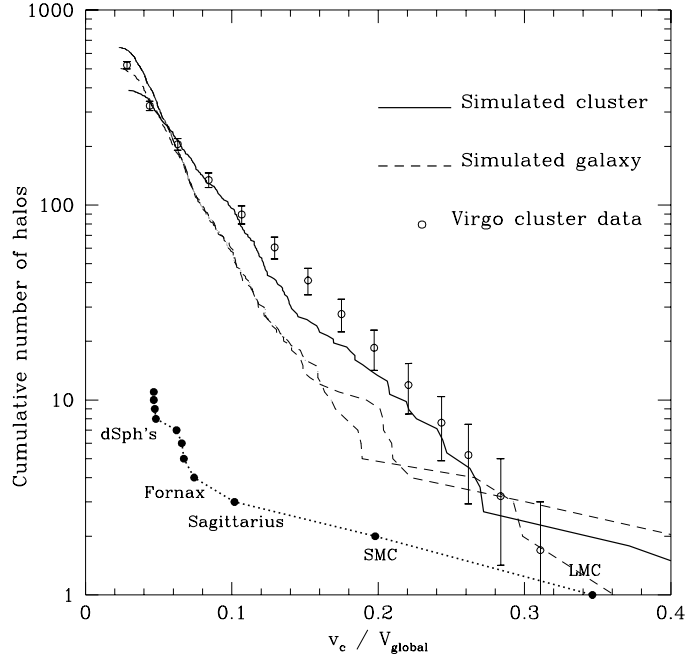}}
\caption[]{Satellite number as a function of circular velocity for the Galaxy
and the Virgo cluster as observed (dots) and as predicted by hierarchical models
 (dashed and solid lines, for the Galaxy and Virgo respectively). From Moore et
 al. (1999).}
\label{counts}
\end{figure}

Could they be the most compact high velocity clouds (CHVCs) observed in large
number around the Galaxy, as suggested by Blitz et al (1999) ? Many observational
campaigns (e.g. Gibson 2001, Simon and Blitz 2002, Hopp, this volume) have     
been performed to search for stars in these systems, under the hypothesis that
if they are the equivalent of the Galactic building blocks they must 
contain  stars. However, no star has ever been found in a HVC,
despite the fact that all these campaigns had sufficient sensitivity to
detect them.

Could most of the satellites be invisible because their SF has been inhibited by
early reionization, as suggested e.g. by MacLow and Ferrara (1999) and Bullock et
al. (2000) ? But then, why haven't visible dwarfs been inhibited too ?

Or, is the satellite overprediction the consequence of having assumed the dark
matter to be cold, while it should actually be assumed to be warm ?

\subsection {Miscellanea}

There are several other aspects to discuss to better understand whether or not  
present-day
dwarf galaxies can be considered the analogues of galactic building blocks.

\noindent {\bf Carbon stars}:
Present-day dwarfs contain many C-stars, while the halos of big galaxies like
the Milky Way don't. If the latter have formed from dwarfs merging, where have
their C-stars gone ? Probably, the only way to avoid the overprediction of halo
C-stars is to assume that the halo itself formed more than 10 Gyr ago, i.e.
before the birth of C-star progenitors (e.g. van den Bergh 1996).

\noindent {\bf Half-light radii of globular clusters}:
The half-light radii of globular clusters in the Galactic halo are
tightly correlated with their Galactocentric distance, a circumstance suggesting
that they are all ruled by a common law. Since the Galactic half-light radii are
all larger than those of globular clusters e.g. in the nearby dwarf galaxy
Fornax, how can our halo clusters come from the accretion of Fornax-like systems
(e.g. van den Bergh 1996) ?

\noindent {\bf Kinematics}: 
Standard CDM models of hierarchical galaxy          
formation have several problems in predicting the kinematic properties of
present day galaxies. One is the disk overheating due to the energy
transferred by the accreted fragments (e.g. Toth and Ostriker 1992, Moore et al.
1999). For instance, Torres et al (2001) suggest that the kinematic properties
of white dwarfs in the Galactic disk are compatible only with small accretion
episodes (i.e. with satellite mass $\leq 4\% $  of the Milky Way mass) occurred
earlier than $\sim$6 Gyr ago, and are definitely inconsistent with more recent
(in the last 6 Gyr) or massive (satellite with mass $\geq 16\% $ of Milky Way
mass) ones.

The so-called {\it angular momentum catastrophe} is probably the worst 
kinematic problem: the dynamical friction of the orbiting gas clumps and the
gravitational torques exerted by non-spherical DM distributions make the angular
momenta predicted by hierarchical scenarios for spiral disks more than 10 times
lower than observed and no convincing way out of this inconsistency has been
found yet (e.g. Navarro and Steinmetz, 2000). 

\section {Discussion}

The topics described above were      
extensively debated during a very lively discussion, with an audience fairly
well balanced between supporters of the {\it hierarchical} and of the           
{\it monolithic} schools of thought. We defined as {\it hierarchical} all the
models (CDM, WDM, etc.) predicting  that galaxies form by successive merging of
lower mass fragments, which inevitably imply that more massive galaxies form
later than lower mass ones, and as {\it monolithic} all the models assuming     
galaxies  of any size to form from the collapse of one or more gas clouds,      
which does not imply any delay in the formation of massive systems.

 We all agreed that some of the age, chemical and dynamical
inconsistencies 
 between the observed properties of present-day dwarf galaxies and
those of big galaxies make dwarfs very unlikely to be the major building blocks
of systems like the Milky Way and normal ellipticals. It was however argued
(by M. Steinmetz) that this is not an argument against hierarchical galaxy      
formation theories, but evidence that present-day dwarfs are simply not the
local counterpart of the building blocks of the theoretical CDM models. In his
opinion the actual building blocks may be completely invisible nowadays and
different from any kind of known systems. This may sound as a fairly {\it ad hoc}
way out of the problem, but we cannot exclude it. It requires, at any
rate, that these unknown systems either don't form stars at all before merging
or form them with the appropriate characteristics. 

One of the most interesting
novelties of this debate was the assertion (e.g. by S. Lamb) that in some
hierarchical galaxy formation models the creation of large galaxies by major
mergers of small fragments occurs when the baryonic matter of the latter is     
still fully gaseous. In this case, most of the problems of published CDM galaxy
formation models would be overcome: the stars of any galaxy would form {\it in
situ} just as in the {\it monolithic} scenario. The whole question would then
concern the timescales for the formation of galaxies of different size and
morphological type. If some of the cold and warm dark matter models already
allow for early formation even of big ellipticals, as stressed by some of the
participants (e.g. R. Dominguez), we may dare foreseeing that, hopefully soon,
the two schools of thought will converge in a compromising scenario, where
both the cosmological and the local properties derived from observations will
be reproduced by theoretical models.

\acknowledgements
I thank Ulrich Hopp for providing the data of Table 1 and for very interesting  
conversations on BCDs. Lara Baldacci and Gisella Clementini kindly provided
their results on NGC 6822 in advance of publication, and Matthew Shetrone made
his figure available in convenient format.
This work has been partially supported by the Italian ASI and MIUR through      
contracts ASI-I/R/35/00 and Cofin--2000.

\end{article}
\end{document}